\documentclass[11pt,letterpaper,notoc,nohyper]{JHEP3}
\pdfoutput=1
\usepackage{graphics}
\usepackage{amsmath}
\usepackage{cite}
\usepackage{colordvi}
\usepackage{epsfig}
\usepackage{latexsym}

\newcommand{\slashchar}[1]{\setbox0=\hbox{$#1$}   
   \dimen0=\wd0                                     
   \setbox1=\hbox{/} \dimen1=\wd1                   
   \ifdim\dimen0>\dimen1                            
      \rlap{\hbox to \dimen0{\hfil/\hfil}}          
      #1                                            
   \else                                            
      \rlap{\hbox to \dimen1{\hfil$#1$\hfil}}       
      /                                             
   \fi}                                             %

\newcommand{\ETmiss}{\slashchar{E}_T}
\newcommand{\ra}{\rightarrow}

\newcommand{\Cp}{\tilde{\chi}^{+}}

\newcommand{\Cpm}{\tilde{\chi}^\pm}

\newcommand{\Cpmone}{\Cpm_1}
\newcommand{\N}{\tilde{\chi}^0}
\newcommand{\None}{\N_1}

\title{\mbox{Supersymmetry with a Chargino NLSP and Gravitino LSP}}
\author{%
Graham D. Kribs$^a$, Adam Martin$^b$  and Tuhin S. Roy$^a$ \\ 
\llap{$^a$} Department of Physics and Institute of Theoretical Science, \\
\hspace*{0.1mm} University of Oregon, Eugene, OR 97403 \\
\llap{$^b$} Department of Physics, Sloane Laboratory, Yale University, 
New Haven, CT 06520 
}

\abstract{
We demonstrate that the lightest chargino can be lighter than 
the lightest neutralino in supersymmetric models with Dirac gaugino
masses as well as within a curious parameter region of the MSSM\@.
Given also a light gravitino, such as from 
low scale supersymmetry breaking, this mass hierarchy leads to 
an unusual signal where every superpartner cascades down to a 
chargino that decays into an on-shell $W$ and a gravitino, 
possibly with a macroscopic chargino track.  
We clearly identify the region of parameters where this signal
can occur.  We find it is generic in the context of the $R$-symmetric 
supersymmetric standard model, whereas it essentially only
occurs in the MSSM when 
${\rm sign}(M_1) \not= {\rm sign}(M_2) = {\rm sign}(\mu)$
and $\tan\beta$ is small.
We briefly comment on the search strategies for this signal 
at the LHC\@.}

\begin{document}

\section{Introduction}

If the fundamental scale of supersymmetry breaking is low,
which can happen with gauge mediation \cite{Giudice:1998bp},
the lightest supersymmetric 
particle (LSP) is the gravitino.  Sparticles produced at colliders
will rapidly decay down to the next-to-lightest sparticle (NLSP)
which then slowly decays into a particle and gravitino.
This decay chain is assured assuming there is no excessively
small kinematic suppression for heavier sparticles to 
decay into the NLSP (see e.g.\ \cite{Ambrosanio:1997bq}).
The identity of the NLSP becomes paramount to determine
the collider signal; many possibilities for the NLSP have 
been considered
\cite{Cabibbo:1981er,Dimopoulos:1996vz,Ambrosanio:1996jn,
Dimopoulos:1996yq,Bagger:1996bt,Ambrosanio:1997rv,Raby:1997bpa}
including the lightest neutralino, the stau, and the gluino.

In this paper, we demonstrate that the NLSP could be a chargino,
leading to a dramatically distinct signal of supersymmetry.
Every superpartner cascades down to a chargino that decays 
into an on-shell $W$ and a gravitino, possibly with a 
macroscopic chargino track.  The final decay 
$\Cpm \ra W^\pm \tilde{G}$ is 2-body, at least for 
$m_{\tilde{G}} < 21$ GeV, due to the LEPII bound on the 
mass of charginos, $m_{\Cpm} > 101$~GeV \cite{Yao:2006px}.

Common lore asserts that the lightest neutralino is always 
lighter than the lightest chargino in the minimal supersymmetric 
standard model (MSSM).  This is certainly true in the bino limit 
$M_1 \ll M_2,\mu,M_Z$, and has been studied and confirmed  
in the wino limit 
\cite{Randall:1998uk,Giudice:1998xp,Cheng:1998hc,Feng:1999fu}
and Higgsino limit \cite{Giudice:1995qk,Drees:1996pk}, 
at least without excessively large radiative corrections.
Generically, radiative corrections to the mass
\emph{difference} between the chargino and neutralino
are small, less than a GeV \cite{Pierce:1996zz}.
The exception is if the 
lightest gauginos are Higgsino-like with significant 
contributions from top and bottom squarks, where 
Ref.~\cite{Giudice:1995qk} found it could be as large as
a few GeV\@.  It is not clear if this parameter region remains viable,
in light of present direct search constraints and electroweak 
precision corrections.  Nevertheless, as we will see, there are
qualitatively distinct regions of \emph{tree-level}
gaugino parameters resulting in a chargino NLSP regardless
of radiative corrections.  This is the focus of the paper.

We find two qualitatively distinct scenarios where the
chargino can be the NLSP\@.  The first, and by far the most significant,
is the minimal $R$-symmetric supersymmetric model (MRSSM).
Generically, this model can have the lightest chargino lighter 
than the lightest neutralino due to the fundamentally different 
neutralino mass matrix that results from the Dirac gaugino masses.  
The second scenario is, remarkably, a curious and relatively
unexplored region of the MSSM parameter space, where
${\rm sign}(M_1) \not= {\rm sign}(M_2) = {\rm sign}(\mu)$
and $\tan\beta$ is small \cite{othersignnote}.
The mass difference 
between the lightest neutralino and the lightest chargino
in the $R$-symmetric scenario can be tens of GeV or more,
whereas it can be up to about 5 GeV (at tree-level) in this 
curious region of the MSSM\@.  

The organization of this paper is as follows.
Sec.~\ref{dirac-sec} reviews Dirac gauginos, the MRSSM model, 
and demonstrates that a chargino is NLSP in a wide region of
parameter space.  Sec.~\ref{mssm-sec} is devoted to identifying
the curious region of the MSSM parameter space where a chargino 
can be the NLSP\@.  In Sec.~\ref{decay-sec} the decay width of the
chargino NLSP into the LSP is calculated.
In Sec.~\ref{pheno-sec} the collider phenomenology of a 
chargino NLSP is discussed.  Finally, in Sec.~\ref{conclusions-sec}
we conclude.

Many analytical results are presented to concretely demonstrate 
that the chargino can be the NLSP in the wino and Higgsino limits
of the MRSSM and the MSSM\@.  This discussion is somewhat technical; 
readers interested in just knowing the parameter space that results 
in a chargino NLSP may go directly to Sec.~\ref{numerical-MRSSM-sec} 
for the MRSSM 
(especially Figs.~\ref{fig:mrssm3} and \ref{fig:mrssm4}), 
and to the latter half of Sec.~\ref{mssm-sec} for the MSSM
(especially Figs.~\ref{fig:mssm5} and \ref{fig:mssm6}).
Readers unfamiliar with Dirac gauginos are encouraged to read
up to the end of Sec.~\ref{mass-matrix-MRSSM}.
Readers interested in just the new signals can skip
directly to Sec.~\ref{pheno-sec}.

\section{Neutralinos and Charginos with Dirac Gaugino Masses}
\label{dirac-sec}

Dirac gaugino masses result when the gaugino is married with
a fermion in the adjoint representation through the operator
\begin{eqnarray}
\int d^2\theta \frac{W'_\alpha}{M} W^{\alpha}_i \Phi_i \; ,
\label{dirac-eq}
\end{eqnarray}
after the spurion $W'_\alpha = D \theta_\alpha$ acquires a $D$-term.
Here $\Phi_i$ are chiral superfields in the adjoint representation
of the SM groups.  $M$ 
represents the messenger scale where these operators are generated.
This possibility has been contemplated for weak scale supersymmetry 
some time ago \cite{Fayet:1978qc,Polchinski:1982an,Hall:1990hq} 
and more recently
\cite{Fox:2002bu,Nelson:2002ca,Chacko:2004mi,Carpenter:2005tz,Antoniadis:2006uj,Hisano:2006mv,Hsieh:2007wq,Kribs:2007ac}.

Gauginos which acquire Dirac masses from Eq.~(\ref{dirac-eq}) 
are not necessarily Dirac fermions once electroweak symmetry
is broken and the gauginos mix with Higgsinos.
Charginos are obviously Dirac fermions, since charginos
carry a conserved $U(1)$ charge, i.e., electric charge.
Neutralinos are Dirac fermions only
if a global $U(1)$ is preserved by all neutralino interactions.
In the minimal $R$-symmetric supersymmetric standard model 
(MRSSM) \cite{Kribs:2007ac}, 
a $U(1)_R$ symmetry is preserved, and thus the neutralinos 
are Dirac fermions.  By contrast, in the Fox-Nelson-Weiner (FNW) 
model \cite{Fox:2002bu}, while gauginos acquire Dirac masses, 
the Higgsinos acquire mass through an ordinary $\mu$-term.
The Higgsino mass violates the $U(1)_R$-symmetry, and thus leads 
to neutralinos that are (pseudo-Dirac) Majorana fermions.
For our purposes, the most illuminating scenario with 
Dirac gaugino masses is the MRSSM\@.  

The remarkable feature of the MRSSM is that it drastically 
ameliorates the supersymmetric flavor problem,
with no excessive contributions to electric dipole moments, 
but with order one squark and slepton mass mixings among 
nearly all flavors \cite{Kribs:2007ac,Blechman:2008gu}.
This is possible for several reasons:
left-right squark and slepton mixing is absent;
the gaugino masses $M$ can be naturally $4\pi/g$ heavier 
than the scalar masses; and several flavor-violating
operators are more suppressed than in the MSSM due to the absence 
of $R$-violating operators.

A low energy model with $U(1)_R$ symmetry can arise, for example, 
if supersymmetry breaking hidden sector preserves $U(1)_R$, which 
happens in a wide class of supersymmetry breaking models
(see e.g., \cite{Intriligator:2006dd}).
Nevertheless, cancellation of the cosmological constant with an 
$R$-violating constant in the superpotential \cite{Bagger:1994hh}
is generally expected to cause $R$-violation to be communicated
from the hidden sector to the MRSSM via anomaly mediation.  
A natural way to minimize 
the size of the $R$-symmetry violation is to take the gravitino 
mass small, such as in a low scale supersymmetry breaking scenario.
It is thus very natural to imagine an $R$-symmetric model
with a light gravitino, making the resulting experimental signals
important to study.

There are several other models that have Dirac gaugino masses
where we find that the chargino NLSP phenomenon can also occur.  
In Appendix~\ref{supersoft-app} we briefly comment on the 
FNW model \cite{Fox:2002bu}, showing that there are specific limits
where the neutralinos become Dirac fermions and the mass matrices
reduce to the ones found in the MRSSM\@.  Hence, our results 
apply to this model as well.

\subsection{Mass Matrices of the MRSSM}
\label{mass-matrix-MRSSM}

In the MRSSM, gaugino masses arise from Eq.~(\ref{dirac-eq})
which generate Dirac masses that pair the gauginos 
$\left( \tilde{g} , \tilde{W} \; \text{and} \; \tilde{B}\right) $ 
with their fermionic partners 
$\left( \psi_{\tilde{g}}, \psi_{\tilde{W}} \; \text{and} \; \psi_{\tilde{B}}\right)$. 
Higgsino masses arise from pairing Higgs superfields
$H_u$ and $H_d$ with partner fields $R_u$ and $R_d$
through a pair of mass terms
\begin{equation}
  \label{eq:mu-term}
  \int \; d^2\theta \; \Big[ \mu_u H_u R_u + \mu_d H_d R_d \Big] \; .
\end{equation}
$R_{d,u}$ transform identically to $H_{u,d}$ under the electroweak
group, except that the $R$-charges are $2$ rather than $0$.
This $R$-charge assignment forbids Yukawa-like couplings of the 
$R$-fields to the matter fields.  Hence, only Higgses acquire 
electroweak symmetry breaking expectation values.  
Upon electroweak symmetry breaking, the electroweak gauginos 
mix with the Higgsinos (just like in the MSSM).
For completeness, all the multiplets in the MRSSM described in
Ref.~\cite{Kribs:2007ac} are listed in Table~\ref{table:fields}
along with their matter and $R$-charges. 
\begin{table}[t]
\centering 
\begin{tabular}{ c | c c c c }
  Fields \quad & $SU(3)_C$ & $SU(2)_W$ & $U(1)_Y$ & $U(1)_R$ \\ \hline 
  $Q$                & $3$        & $2$ & $\frac{1}{6}$  & $1$   \\
  $U$                & $\bar{3}$  & $1$ & -$\frac{2}{3}$ & $1$   \\
  $D$                & $\bar{3}$  & $1$ & $\frac{1}{3}$  & $1$   \\
  $L$                & $1$        & $2$ & -$\frac{1}{2}$ & $1$   \\
  $E$                & $1$        & $1$ & $1$            & $1$   \\
  $\Phi_{\tilde{B}}$ & $1$        & $1$ & $0$            & $0$   \\
  $\Phi_{\tilde{W}}$ & $1$        & $3$ & $0$            & $0$   \\
  $\Phi_{\tilde{g}}$ & $8$        & $1$ & $0$            & $0$   \\
  $H_u$              & $1$        & $2$ & $\frac{1}{2}$  & $0$   \\
  $H_d$              & $1$        & $2$ & -$\frac{1}{2}$ & $0$   \\
  $R_u$              & $1$        & $2$ & -$\frac{1}{2}$ & $2$   \\
  $R_d$              & $1$        & $2$ & $\frac{1}{2}$  & $2$   \\
\hline
\end{tabular}
\caption{Gauge and $R$-charges of all chiral supermultiplets in the MRSSM\@.}
\label{table:fields}
\end{table}

Let us first investigate the neutralino mass matrix in the MRSSM\@.  
The $R$-charges determine which neutralinos mix with each 
other and provide a guiding principle to determine the 
gauge-eigenstate basis.  The vector 
$N_{+} \equiv \left( \tilde{W}_3, \tilde{B}, 
                     \tilde{R}_u^0, \tilde{R}_d^0 \right)$
carries $R$-charge $+1$, while the vector 
$N_{-} \equiv \left( \psi^0_{\tilde{W}}, \psi_{\tilde{B}}, 
                     \tilde{H}_u^0, \tilde{H}_d^0 \right)$ 
carries a $R$-charge $-1$.  Field rotations do not mix up fields 
with different $R$-charges.  Hence, just like with charginos in the MSSM, 
two independent rotation matrices are required to diagonalize 
the mass matrix.

The Lagrangian for neutralino masses can be written in the 
gauge-eigenstate basis as
\begin{equation}
\mathcal{L}_\text{neutralino mass} = \; \; N_{+}^\text{T} \;  M_{\tilde{N}} \; N_{-}  
       \;\; + \;\; \text{c.c.}      \; ,  
\end{equation}
where the neutralino mass matrix is given as 
\begin{equation}
\label{eq:mrssm-nut-mass}
 M_{\tilde{N}}  = 
\left[ \begin{array}{cccc}  
      M_2 & 0 & - g  v_u/\sqrt{2} &   g  v_d / \sqrt{2}  \\ 
      0 & M_1 &   g' v_u/\sqrt{2} & - g' v_d / \sqrt{2}  \\ 
      - \lambda_u  v_u/\sqrt{2}   &   \lambda_u' v_u/\sqrt{2} & \mu_u & 0 \\ 
        \lambda_d  v_d / \sqrt{2} & - \lambda_d' v_d / \sqrt{2} & 0 & \mu_d 
       \end{array} \right]   \; .
\end{equation}
Here $\langle H_u \rangle \equiv v_u$ and $\langle H_d \rangle \equiv v_d$
with $v_u^2 + v_d^2 = v^2/2 \simeq (174 \; {\rm GeV})^2$.
Notice the apparently unusual location of the $\mu$-terms 
in Eq.~(\ref{eq:mrssm-nut-mass}) is a direct result of the
Dirac nature of the neutralino mass matrix.
The physical mass-squareds
are given by eigenvalues of $M_{\tilde{N}} M_{\tilde{N}}^\text{T}$.
There are four new parameters $\lambda_u, \lambda_d, \lambda_u'$ 
and $\lambda_d'$ that arise from the superpotential terms
\begin{equation}
\label{eq:lambda-term}
\int \; d^2\theta \; 
 \Big[ H_u \left( \lambda_u \Phi_{\tilde{W}} + \lambda_u' \Phi_{\tilde{B}} \right) R_u +
       H_d \left( \lambda_d \Phi_{\tilde{W}} + \lambda_d' \Phi_{\tilde{B}} \right) R_d   \Big] \; .
\end{equation}
These couplings are unnecessary to the structure of the MRSSM
but are nevertheless allowed under all of the charge assignments. 
Various checks have been performed on Eq.~(\ref{eq:mrssm-nut-mass}) 
to verify that every entry in this mass matrix is correct, 
see Appendix~\ref{checks-app}. 

For charginos, the mass matrix is even simpler because 
of the conservation of electromagnetic charge as well as $R$-charge.  
In the MRSSM there are eight
two-component fermions from the winos, Higgsinos, and $R$-fields.
Based on the $R$-charges and the electromagnetic charges, 
the eight fermions can be grouped into following four different 
classes which do not mix among themselves:

\begin{equation}
  \begin{array}{c | c | c }
    \text{charges} & Q = +1  & Q = -1 \\
         \hline
    R = +1 & \; \; \chi_{++} \equiv \left( \tilde{W}^{+} , \tilde{R}_d^{+} \right) \;
           & \; \; \chi_{+-} \equiv \left( \tilde{W}^{-} , \tilde{R}_u^{-} \right) \\
         \hline 
    R = -1 & \; \; \chi_{-+} \equiv \left( \psi^{+}_{\tilde{W}}, \tilde{H}_u^{+} \right) \; 
           & \; \; \chi_{--} \equiv \left( \psi^{-}_{\tilde{W}} , \tilde{H}_d^{-} \right)  \; \\
  \end{array}
  \label{eq:chargino-basis}
\end{equation}
These charges imply that the charginos pair up as
\begin{equation}
  \label{eq:chargino-mass}
  \mathcal{L}_\text{chargino mass} = \quad 
     \chi_{++}^\text{T}  \;  M_{\chi_1} \; \chi_{--} \;\; + \;\; 
     \chi_{-+}^\text{T}  \;  M_{\chi_2} \; \chi_{+-} \;\; + \;\;
     \text{c.c.}      \; ,    
\end{equation}
where the chargino mass matrices in our basis are 
\begin{equation}
\label{eq:mrssm-chr-mass}
  M_{\chi_1} \; = \; 
       \left[ \begin{array}{cc}
                 M_2    & \lambda_d v_d  \\
                 g v_d  &  \mu_d
               \end{array} \right]   \quad \text{and} \quad
  M_{\chi_2} \; = \; 
       \left[ \begin{array}{cc}
                 M_2 &  \lambda_u v_u \\
                 g v_u &  \mu_u
               \end{array} \right] \; .   
\end{equation}
Since each of these matrices are diagonalized by bi-unitary transformations, 
four independent rotations are needed for the four vectors listed in 
Eq.~(\ref{eq:chargino-basis}) in order to diagonalize the mass matrices.

We are now ready to calculate the mass eigenstates under 
various assumptions about the parameters of the MRSSM\@.  
One qualitative difference from the MSSM is that the Dirac nature 
of the gauginos allows us to take the gaugino masses $M_1,M_2$ 
and Higgsino masses $\mu_u,\mu_d$ real and positive 
(rotating phases into the holomorphic masses for the scalar
adjoints and the $\lambda$ parameters) \cite{Kribs:2007ac}.
Examining the gaugino mass matrices, we explore several limits 
where we can analytically demonstrate that the chargino 
is the NLSP and obtain a good estimate of the mass difference.
Our first treatment is to take $\lambda_{u,d} = \lambda'_{u,d} = 0$.  
This is motivated in part to simplify our analysis, but also to 
emphasize that nonzero values of these couplings are \emph{not} 
necessary to obtain a chargino lighter than the lightest neutralino. 
Further simplifications can result by taking the large $\tan \beta$ limit
and the large $M_1$ limit. In Sec.~\ref{lambda-MRSSM-sec} we 
generalize and expand the discussion, 
while still working at tree-level.
As we will see, the lightest chargino can be \emph{significantly} 
lighter than the lightest neutralino when the $\lambda$ couplings
are present with ${\cal O}(g)$ values.

\subsection{Simplified MRSSM: Large $M_1$ with $\lambda = \lambda' = 0$}
\label{large-MRSSM-sec}

The gaugino mass matrices,
Eqs.~(\ref{eq:mrssm-nut-mass}) and (\ref{eq:mrssm-chr-mass}),
simplify in the limit $\lambda_{u,d} = \lambda'_{u,d} = 0$.
Here we will also take $\mu_u = \mu_d = \mu$, which will prove 
extremely convenient in our analytic analysis in this section.
Equal Higgsino masses will also allow us to illustrate the contrast 
between the MSSM and the MRSSM\@.  We are obviously 
not interested in the case where the bino is the lightest gaugino,
hence we take large $M_1$, consistent with the motivations of
Ref.~\cite{Kribs:2007ac}.  Integrating out the bino and taking 
$\tan{\beta} \gg 1$, 
the neutralino and chargino mass matrices are given by
\begin{equation}
  \label{eq:mass-limit-rssm1}
M_{\tilde{N}} = 
  \left[ \begin{array}{c c c}
                 M_2 &  -  M_W & 0 \\
                 0   &  \mu  & 0 \\
                 0   &  0    & \mu
               \end{array} \right]  
\quad \quad \text{and} \quad \quad
M_{\chi_2} = 
  \left[ \begin{array}{c c}
                 M_2 &  \sqrt{2} M_W  \\
                 0   &  \mu  
               \end{array} \right]  \; .  
\end{equation}
For charginos, in the case $\tan\beta > 1$, we need only consider 
$M_{\chi_2}$ in order to find the lightest chargino mass.

The block diagonal form of $M_{\tilde{N}}$ in Eq.~(\ref{eq:mass-limit-rssm1}) 
shows that a pair of neutral Higgsinos acquire a (Dirac) mass $\mu$ 
and do not mix with the other neutralinos.  This allows us to
further simplify the neutralino mass matrix down to just the
upper $2\times 2$ block.  Here is the key analytical observation:  
The upper $2\times 2$ block of $M_{\tilde{N}}$ is identical to 
$M_{\chi_2}$, except that the off-diagonal element is 
\emph{smaller} for the neutralino mass matrix.  
Simple $2\times 2$ diagonalization clearly shows that the 
lightest chargino is lighter than the lightest neutralino in 
{\em both} the Higgsino limit ($\mu < M_1, M_2$) and the wino limit 
($M_2 < \mu, M_1$), 

\begin{equation}
  \begin{split}
\text{Higgsino limit:}& \quad
    \Delta_{+} = 
      - \; \frac{ \mu M_W^2 }{2 M_2^2}
       + \mathcal{O}\left( \frac{1}{M_2^4} \right) \; , \\
\text{wino limit:}& \quad
    \Delta_{+}
     =  - \; \frac{M_2 M_W^2}{2 \mu^2}
       + \mathcal{O}\left( \frac{1}{\mu^4} \right) \; .
  \end{split}
\end{equation}   
where $\Delta_+ \equiv m_{\Cpmone} - m_{\None}$.

It is well known that the one-loop electromagnetic radiative
correction increases the chargino mass
\cite{Randall:1998uk,Giudice:1998xp,Cheng:1998hc,Feng:1999fu}.
But, since the size of the mass difference shown above is 
not parametrically small
compared to the loop contribution, the tree-level splitting 
can easily dominate so long as one of the diagonal elements 
does not far exceed the other.

\begin{figure}[t]
\centering
\begin{minipage}[c]{0.45\linewidth}
   \centering
   \includegraphics[width = 2.9 in]{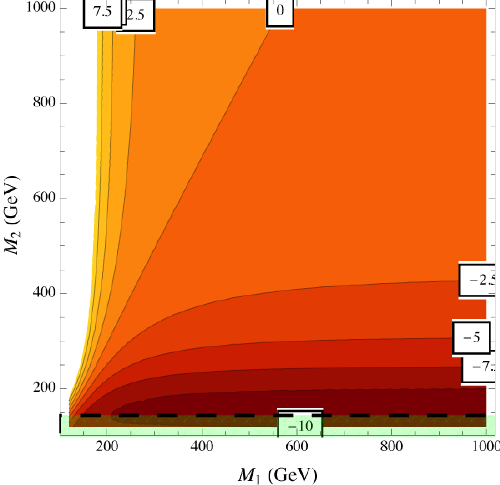} 
   \caption{Contours of $\Delta_{+}$(GeV) in the simplified 
      MRSSM at $\tan\beta=10$ and $\mu = 150$~GeV\@.}
   \label{fig:mrssm1}
\end{minipage}
\hspace{0.5 cm}
\begin{minipage}[c]{0.45\linewidth}
   \centering
   \includegraphics[width = 2.9 in]{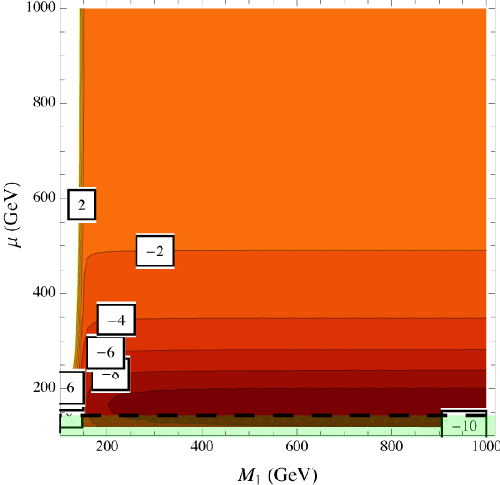}
   \caption{Contours of $\Delta_{+}$(GeV) in the simplified 
      MRSSM at $\tan\beta=10$ and $M_2 = 150$~GeV\@.}
   \label{fig:mrssm2}
\end{minipage}  
\end{figure}

The analytical results are strikingly confirmed by examining
a larger region of the MRSSM parameter space numerically.
In Figs.~\ref{fig:mrssm1} and \ref{fig:mrssm2} we plot the
contours of $\Delta_{+}$, the difference of 
the lightest chargino mass to the lightest neutralino mass. 
Fig.~\ref{fig:mrssm1} explores the mostly-Higgsino limit and was 
generated by holding $\mu = 150$~GeV and $\tan\beta = 10$.
Similarly, Fig.~\ref{fig:mrssm2} explores the mostly-wino limit 
and was generated holding $M_2 = 150$~GeV and $\tan\beta = 10$.  
The regions under the dashed lines in these Figures result 
in $m_{\Cpmone} < 101$~GeV at tree level and thus will be ignored
from further consideration. 

In these Figures, it is clear that a chargino is the lightest 
gaugino in the regions with $\Delta_{+} < 0$.  This occurs 
throughout the tree-level parameter space of the wino limit, shown 
in Fig.~\ref{fig:mrssm2}.  Note that we have not included
radiative corrections in these numerical results because
a full calculation requires knowing the full spectrum 
of the model.  Nevertheless, it is 
clear that in a wide range of parameter space, the tree-level 
mass difference is much larger than are expected from radiative
corrections, demonstrating that the chargino can indeed be the NLSP\@.

\subsection{Enlarging the parameter space:  nonzero $\lambda$ couplings}
\label{lambda-MRSSM-sec}
The full parameter space of the MRSSM is much larger than the MSSM\@.
To keep things manageable  
we consider the following simplifications:
$\lambda_u = \lambda_d \equiv \lambda$ and 
$\lambda'_u =\lambda'_d \equiv \lambda'$. 
Moreover, it will be useful to define the parameters 
$M_\lambda =  \lambda v/2 $ and 
$\tan \theta_\lambda = \lambda'/\lambda$, analogous to the 
electroweak parameters $M_W = g v/2$ and 
$\tan\theta_W = g'/g$. 
We use perturbation techniques on this mass-squared matrix to find 
approximate analytical expressions for neutralino masses.

\subsubsection{Equal Higgsino masses $\mu_u = \mu_d = \mu$} 

In this case, the neutralino mass matrix simplifies
in a very interesting way:
\begin{equation}
 M_{\tilde{N}} =
\left[ \begin{array}{cccc}  
M_2 & 0   & -\sin \beta  \; M_W                & \cos \beta  \; M_W \\ 
0   & M_1 & \sin \beta \; \tan \theta_W \; M_W 
                             & -\cos \beta \; \tan \theta_W \; M_W \\ 
-\sin \beta  \; M_\lambda   & \sin \beta \; \tan \theta_\lambda \; M_\lambda & 
                                    \mu & 0 \\ 
      \cos \beta  \; M_\lambda & -\cos \beta \; \tan \theta_\lambda \; M_\lambda & 
                                        0 & \mu 
       \end{array} \right]   \; .
\end{equation}
The lower $2 \times 2$ block is proportional to the identity matrix, 
which implies that an orthogonal transformation involving only the 
last two columns and rows will leave the lower $2\times 2$ block invariant. 
Moreover, the ratio of the third and the fourth elements 
in the first two rows and as well as in the first two columns is
$\tan \beta$.  Therefore a rotation of the last two rows and
columns by an angle $\beta$ will make the third element of the first
two rows and columns vanish simultaneously while leaving
the lower $2 \times 2$ block unchanged.  In this new basis, we then 
have zeroes everywhere in the third row and column except
at the diagonal position.  Thus we have managed to decouple a pair 
of neutralinos of mass $\mu$ from the other eigenstates.

The chargino mass-squared are similarly found after diagonalizing
$M_{\chi_1} M_{\chi_1}^\text{T}$ and $M_{\chi_2} M_{\chi_2}^\text{T}$. 
These rank two matrices are straightforward to evaluate.
For the purpose of comparing with the neutralino masses, however,
we expand the eigenvalues in various limits. 
When $\tan \beta > 1$, the lightest chargino mass is found from 
$M_{\chi_2}$.  For the mass difference, we find
\begin{itemize}
\item {\bf Higgsino limit:  $M_1,M_2 > \mu$}
\begin{equation}
\label{eq:mrssm-mass-lim1}
    \Delta_{+} = - \;  M_\lambda M_W \left(
        \frac{2  \sin^2 \beta - 1}{M_2} - 
         \frac{\tan\theta_W \tan\theta_\lambda}{M_1} \right) + 
        \mathcal{O}\left( \frac{1}{M_2^2}\; , \frac{1}{M_1^2} \right)  \; ,
\end{equation}  
\item {\bf wino limit: $ M_1,\mu > M_2$}
\begin{equation}
\label{eq:mrssm-mass-lim2}
   \Delta_{+} = - \;    
       \left( 2 \sin^2\beta - 1 \right) 
              \frac{  M_\lambda M_W}{\mu}  + 
        \mathcal{O}\left( \frac{1}{\mu^2} \; , \frac{1}{M_1^2} \right)  \; .
\end{equation}

\end{itemize}

To leading order in this expansion, a chargino is clearly the lightest 
gaugino in the wino limit. In the Higgsino limit, the ratio 
of $M_2$ and $M_1$ is important.  When 
$M_1/M_2 > \tan\theta_W \tan\theta_\lambda / (2 \sin^2 \beta - 1)$, we again 
find a lighter chargino.   

\subsubsection{One Higgsino Heavy:  $\mu_d \gg M_1, M_2$ and $\mu_u = \mu$}

Another limit which can be analyzed analytically is when one Higgsino is much 
heavier than the other mass parameters. Taking $\mu_d \gg M_1, M_2$,  we can 
immediately integrate out down-type Higgsinos 
resulting in a $3 \times 3$ neutralino matrix.  For charginos,
we need to take into account only $M_{\chi_2}$. Again using perturbation 
techniques, we find

\begin{itemize}

\item {\bf  Higgsino limit:  $M_1,M_2 > \mu$}

  \begin{equation}
    \label{eq:mrssm-mass-lim3}
     \Delta_{+} = - \; 
         M_\lambda M_W\; \sin^2 \beta 
            \left(  \frac{1}{M_2} - 
         \frac{\tan\theta_W \tan\theta_\lambda}{M_1} \right) + 
        \mathcal{O}\left( \frac{1}{M_2^2}\; , \frac{1}{M_1^2} \right)  \, ,
  \end{equation}
\item {\bf wino limit: $ M_1,\mu > M_2$}
\begin{equation}
  \label{eq:mrssm-mass-lim4}
    \Delta_{+} = - \;  
               \sin^2\beta  \;
              \frac{  M_\lambda M_W}{\mu}  + 
        \mathcal{O}\left( \frac{1}{\mu^2} \; , \frac{1}{M_1^2} \right)  \; .
\end{equation}
\end{itemize}
The mass differences calculated in this case are quite similar 
to the case of equal Higgsino masses, except for the dependency on
$\sin^2 \beta$.  In fact, in the large $\tan \beta$ limit, 
Eqs.~(\ref{eq:mrssm-mass-lim3}) and (\ref{eq:mrssm-mass-lim4}) 
reduce to Eqs.~(\ref{eq:mrssm-mass-lim1}) and (\ref{eq:mrssm-mass-lim2}) 
respectively.  A careful gaze reveals that, at sizeable $\tan \beta$, 
the relevant portion of the gaugino mass matrices are identical 
whether one considers the equal Higgsino case or one Higgsino heavy case.

\subsection{Numerical results}
\label{numerical-MRSSM-sec}

Having demonstrated analytically that the chargino can be the 
NLSP in several limits, we now turn to analyzing a larger
region of the MRSSM parameter space numerically.
The neutralino and chargino masses are determined by 
nine parameters:  $M_1$, $M_2$, $\mu_u$, $\mu_d$, $\tan\beta$,
$\lambda_u$, $\lambda_d$, $\lambda_u'$, and $\lambda_d'$.
As it is too cumbersome to do a complete scan, we restrict to the 
simplifications introduced in the previous subsection:
$\lambda_u = \lambda_d = \lambda$ and $\lambda_u' = \lambda_d' = \lambda'$.
As before, we also trade the parameters $\lambda$ and $\lambda'$ for the 
mass parameters $M_\lambda$ and the angle $\theta_\lambda$. 

In the first part of this subsection we keep $M_\lambda$ and 
$\tan \theta_\lambda$ fixed and scan the rest of the parameter space 
in order to understand the variation of $\Delta_{+}$. Later we will 
choose a particular point in the $M_1$, $M_2$, $\mu_u$, $\mu_d$ and  
$\tan\beta$ space,  and see the dependence of $\Delta_{+}$ on our
choice of  $M_\lambda$ and $\tan \theta_\lambda$. 

Note that given renormalization group evolution of the superpotential 
parameters $\lambda$ and $\lambda'$, evidently  a natural choice is 
$\tan\theta_\lambda = \tan \theta_W$. We also use $M_\lambda = M_W$ to 
begin our discussion.   
\begin{figure}[t]
\centering
\begin{minipage}[c]{0.45\linewidth}
   \centering
   \includegraphics[width = 2.9 in]{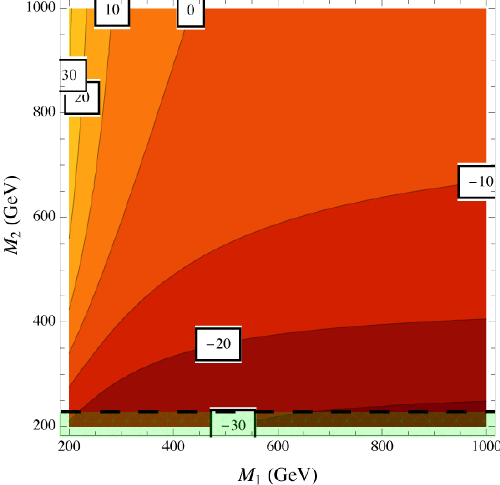} 
   \caption{Contours of $\Delta_{+}$(GeV) in the MRSSM at $\tan\beta=10$ 
and $\mu = 200$~GeV in the equal Higgsino mass limit.}
   \label{fig:mrssm3}
\end{minipage}
\hspace{0.5 cm}
\begin{minipage}[c]{0.45\linewidth}
   \centering
   \includegraphics[width = 2.9 in]{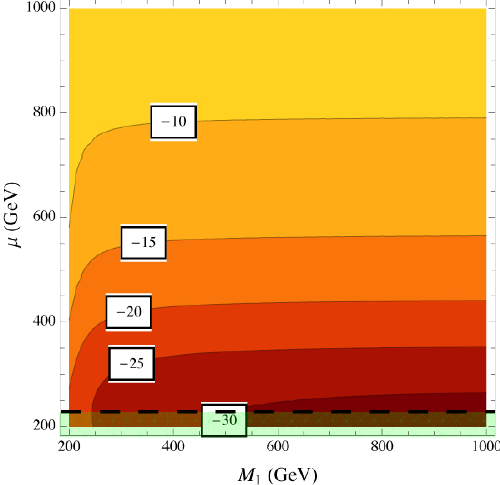}
   \caption{Contours of $\Delta_{+}$(GeV) in the MRSSM at 
$\tan\beta=10$ and $M_2 = 200$~GeV in the equal Higgsino mass limit.}
   \label{fig:mrssm4}
\end{minipage}  
\end{figure}

First, consider the case of equal Higgsino masses with $\tan\beta =10$.  
(The one heavy Higgsino scenario, as we described before, is similar 
to the equal Higgsino scenario at sizeable $\tan\beta$).
In Fig.~\ref{fig:mrssm3} we hold $\mu$ constant and vary 
$M_2$ and $M_1$.  The shape of the contours are the same as in 
Fig.~\ref{fig:mrssm1}, where we neglected the  ``$\lambda$'' couplings.
The striking difference is that now the mass difference ($\Delta_{+}$) 
can be as large as $-30$~GeV\@.
In fact, this is why we chose $\mu = 200$~GeV, so that the 
lightest chargino remains above the LEPII limit.  
In Fig.~\ref{fig:mrssm4}, the same analysis is repeated in the 
wino limit, with $M_2 = 200$~GeV, where again we see that 
mass difference $\Delta_{+}$ is negative and can be up to
$-30$ GeV\@.  

\begin{figure}[h!]
\centering
\begin{minipage}[c]{0.45\linewidth}
   \centering
   \includegraphics[width=2.9in]{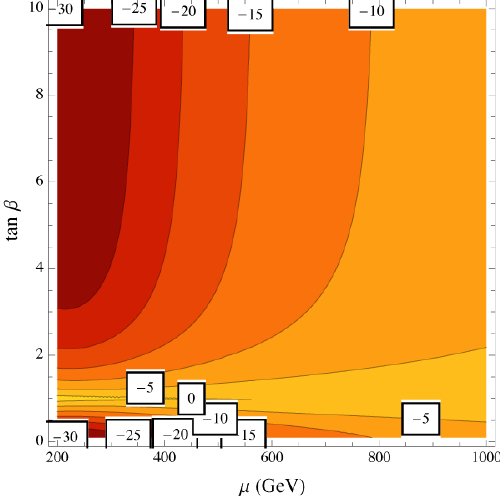} 
   \caption{Contours of $\Delta_{+}$(GeV) in the MRSSM at $M_1 = 500$~GeV and 
            $M_2 = 200$~GeV in the equal Higgsino limit.}
   \label{fig:mrssm5}
\end{minipage}
\hspace{0.5 cm}
\begin{minipage}[c]{0.45\linewidth}
   \centering
   \includegraphics[width=2.9in]{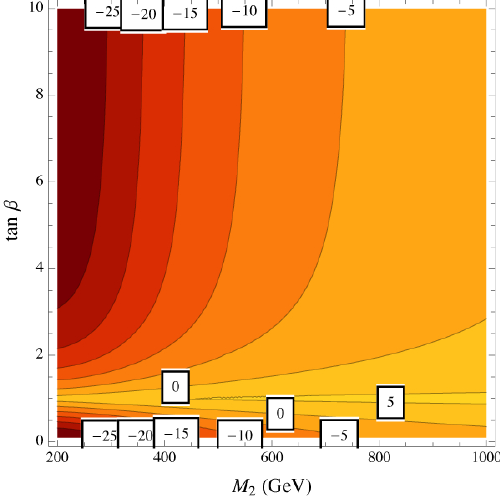}
   \caption{Contours of $\Delta_{+}$(GeV) in the MRSSM at $M_1 = 500$~GeV and 
            $\mu = 200$~GeV in the equal Higgsino limit.}
   \label{fig:mrssm6}
\end{minipage}  
\end{figure}

It is interesting to investigate the dependence of $\Delta_{+}$ on 
$\tan\beta$ in the wino and Higgsino limits.
In Fig.~\ref{fig:mrssm5}, we take $M_1 = 500$~GeV, $M_2  = 200$~GeV,
and vary $\mu$ and $\tan\beta$.  In Fig.~\ref{fig:mrssm6}, 
we take $M_1 = 500$~GeV, $\mu = 200$~GeV, and vary $M_2$ and
$\tan\beta$.  Both the Figures show similar features. 
For a given set of mass parameters, the contours are largely insensitive
to the value of $\tan\beta$ as long as $\tan\beta$ is sizeable.
Close to $\tan\beta = 1$ the contours change rapidly.  At 
$\tan \beta = 1$ we find $\Delta_{+} \geq 0$ in both the wino and Higgsino
cases.  The case $\tan\beta < 1$ is symmetric with respect to
$\tan\beta > 1$ upon taking $\tan \beta \ra 1/\tan\beta$.

We can also investigate how $M_\lambda$ and $\tan\theta_\lambda$
affects the chargino/neutralino mass hierarchy. 
This is most easily done by considering a specific point in the 
$M_1$, $M_2$, $\mu$ and $\tan\beta$ space where $\Delta_{+}$ is large.
From Figs.~\ref{fig:mrssm3} and \ref{fig:mrssm4}, 
we find large (negative) $\Delta_{+}$ when $M_2 \sim \mu$. 
In addition, $\Delta_{+}$ is almost independent of $M_1$ 
for large enough $M_1$.  Finally, from Figs.~\ref{fig:mrssm5} and 
\ref{fig:mrssm6} we see that large $\Delta_+$ occurs at large 
$\tan\beta$.

\begin{figure}[h!]
\centering
\includegraphics[width=0.8\textwidth, height = 3.0in]{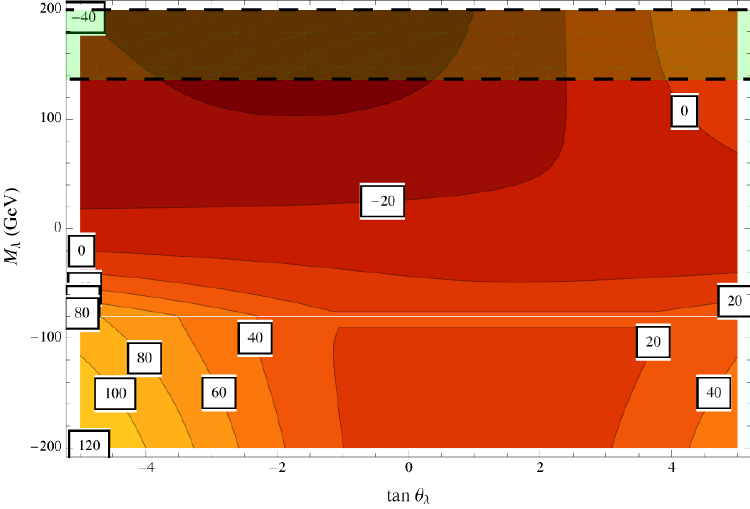}
\caption{Contours of $\Delta_{+}$(GeV) in the $\tan\theta_\lambda-M_\lambda$
plane for $M_1=600\ {\rm GeV},\  \tan{\beta} = 10$ and 
$\mu = M_2 = 250\ {\rm GeV}$.}
\label{fig:mrssm7}
\end{figure}

In Fig.~\ref{fig:mrssm7}, the variation of $\Delta_{+}$ in the 
$M_\lambda - \tan\theta_\lambda$ plane is shown.
We take $M_1 = 600$~GeV and $\tan\beta = 10$ with
$\mu = M_2 = 250\ {\rm GeV}$.  We do not vary $M_\lambda$ beyond 
$\pm 200\ {\rm GeV}$, since even at this value $\lambda$ already 
exceeds 1.  The Figure clearly shows that we get the largest 
(negative) splitting when $M_\lambda$ is positive and 
$\tan\theta_\lambda$ is negative, up to 
$\Delta_{+} \sim - \mathcal{O}(50\ {\rm GeV})$.
The shaded area above the dashed line is excluded since it results in 
a chargino with mass $m_{\Cpm_1} < 101$ GeV\@.  
If $\mu = M_2$ is reduced, $\Delta_+$ becomes larger (negative).  
On the other hand, the upper bound on $M_\lambda$ from the bound
on the chargino mass also decreases rapidly. 
For larger values of $\mu = M_2$, the upper bound on $M_\lambda$ 
increases and effectively one can also find a larger splitting.

\subsection{Summary and Comments}

From various analytical results and numerical figures
it is clear that a chargino can be significantly lighter than the 
lightest neutralino.  The mass difference can far exceed the size
of radiative corrections to the gaugino masses.
\begin{itemize}
\item  The expressions for the mass differences between the lightest neutralino 
and the lightest chargino in Eq.~(\ref{eq:mrssm-mass-lim1}) and in 
Eq.~(\ref{eq:mrssm-mass-lim2}) were obtained for $\tan \beta > 1$. The case
$\tan \beta < 1$ can be obtained by substituting 
$\sin \beta \leftrightarrow \cos \beta$.
\item  The expressions for $\Delta_{+}$ are given in 
Eq.~(\ref{eq:mrssm-mass-lim3}) and in Eq.~(\ref{eq:mrssm-mass-lim4})
when $\mu_d$ is large. An alternate
limit where $\mu_u$ is greater than all other relevant masses may be 
found by substituting
$\sin \beta \leftrightarrow \cos \beta$ 
in Eq.~(\ref{eq:mrssm-mass-lim3}) and in Eq.~(\ref{eq:mrssm-mass-lim4}).  
\item Given the results from the various limits, combined with 
Figs.~\ref{fig:mrssm5} and \ref{fig:mrssm6}, we find that at 
$\tan \beta = 1$, a neutralino is always the lightest gaugino.
\item Finally, in the absence of $\lambda$ couplings, the mass
difference $\Delta_+$ can be obtained in the Higgsino limit 
\emph{without} decoupling $M_1$.  Starting with Eqs.~(\ref{eq:mrssm-mass-lim1}) 
and (\ref{eq:mrssm-mass-lim3}), hold $\tan\theta_\lambda$ finite, 
while independently taking $M_\lambda \rightarrow 0$.  We obtain
\begin{equation}
   \begin{split}
       \mu_u = \mu_d = \mu   \quad & \rightarrow \quad 
             \Delta_{+} = -\; \mu \;  M_W^2 
            \left(  \frac{2 \sin^2 \beta - 1}{M_2^2} - 
         \frac{\tan^2\theta_W}{M_1^2} \right) \; ,
       \\
        \text{Large }\mu_d   \quad & \rightarrow \quad 
               \Delta_{+} = -\; \sin^2\beta \;  \mu \;  M_W^2 
            \left(  \frac{1}{M_2^2} - 
         \frac{\tan^2\theta_W}{M_1^2} \right) \; .
   \end{split}
\end{equation}
These expressions demonstrate the chargino can be lighter than the neutralino
in the MRSSM without $\lambda$ couplings.
\end{itemize} 

\section{Neutralino and Chargino Masses in the MSSM}
\label{mssm-sec}

We now turn to studying the neutralino and chargino masses in the MSSM\@.
The neutralino mass matrix in the MSSM is rank four, and although exact 
analytical expressions for the eigenvalues exist \cite{Barger:1993gh},
they are not particularly transparent. 
We instead consider several well-known limits where the mass 
difference between the chargino and neutralino can be easily 
calculated analytically. Later in this section we generalize our results 
using numerical calculations.  We allow the mass parameters 
$M_1, M_2, \mu$ to have arbitrary sign, 
though without loss of generality we can take
$M_2 > 0$.  We do not consider arbitrary phases,
since they are severely constrained in the MSSM from the absence
of electric dipole moments \cite{Pospelov:2005pr}.
The neutralino gauge eigenstates include a bino with mass $M_1$, hence in 
the small $M_1$ limit a mostly-bino neutralino will always 
be the lightest gaugino. The nontrivial cases of interest to us will occur 
when $M_1$ is not the smallest parameter in the mass matrix.

The two interesting limits that could have a chargino NLSP are 
the Higgsino limit and the wino limit. In the Higgsino limit
$M_1, M_2 > \mu, M_W$, calculations of the masses of light gauginos  
including radiative corrections can be found 
in \cite{Giudice:1995qk,Drees:1996pk}.  The mass difference
between the lightest chargino and lightest neutralino is
\begin{equation}
   \label{eq:mssm-split-lim-mu}
  \Delta_{+} = \Bigg[ \left(  \tan^2\theta_W \frac{M_2}{M_1} +1
      \right)  + \left( \tan^2\theta_W \frac{M_2}{M_1} - 1 \right)
      \frac{\mu}{| \mu |} \sin{2 \beta} \Bigg] \frac{M_W^2}{2 M_2} +
   \mathcal{O}\left( \frac{1}{M_2^2} \right) \; ,
\end{equation}
The neutralino-chargino mass splitting grows as $M_2$ is reduced.

In the wino limit $M_2 < M_1,\mu$, the lightest neutralino
and the lightest chargino are highly degenerate. 
The mass difference was calculated including one-loop effects in 
\cite{Randall:1998uk,Giudice:1998xp,Cheng:1998hc,Feng:1999fu}.
The tree-level splitting can be obtained by expanding in powers 
of $1/\mu$, with the leading order splitting occurring at 
$\mathcal{O}\left( 1/\mu^2\right)$:   
\begin{equation}
    \label{eq:mssm-split-lim-m2}
  \Delta_{+} = \frac{M_W^2}{\mu^2} \frac{M_W^2}{M_1 - M_2} \tan^2 \theta_W
    \sin^2 2\beta  + \mathcal{O}\left(\frac{1}{\mu^3}\right)  \; .
\end{equation}
One loop corrections 
to Eq.~(\ref{eq:mssm-split-lim-m2}) are positive and typically small,
of order 0.1 GeV \cite{Feng:1999fu}. 
\begin{figure}[t]
\centering
\begin{minipage}[c]{0.45\linewidth}
   \centering
   \includegraphics[width = 2.9 in]{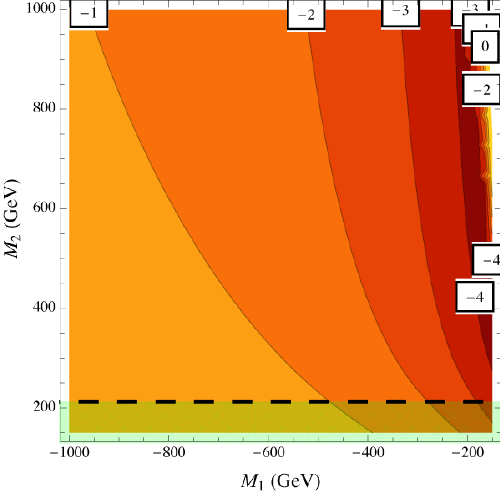} 
   \caption{Contours of $\Delta_{+}$(GeV) in the MSSM at $\tan\beta=2$ and 
            $\mu = 150$~GeV\@.}
   \label{fig:mssm3}
\end{minipage}
\hspace{0.5 cm}
\begin{minipage}[c]{0.45\linewidth}
   \centering
   \includegraphics[width = 2.9 in]{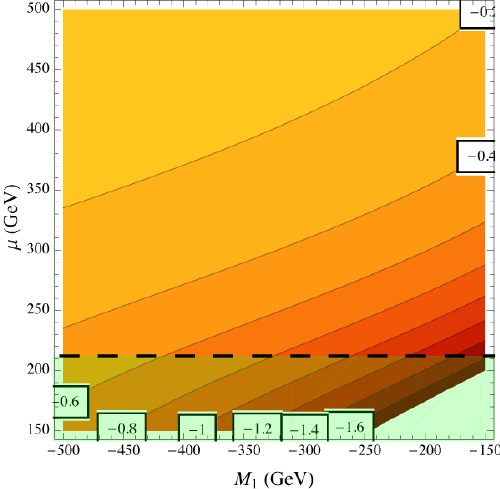}
   \caption{Contours of $\Delta_{+}$(GeV) in the MSSM at $\tan\beta=2$ and 
            $M_2 = 150$~GeV\@.}
   \label{fig:mssm4}
\end{minipage}  
\end{figure}

In the well-known case ${\rm sign}(M_1) = {\rm sign}(M_2)$,
it is evident from the leading terms in Eqs.~(\ref{eq:mssm-split-lim-mu})
and (\ref{eq:mssm-split-lim-m2})
that $m_{\Cpmone} > m_{\None}$ for any sign of $\mu$ in either 
the Higgsino limit or wino limit.  The lightest neutralino persists as the
lightest gaugino throughout the MSSM parameter space with
${\rm sign}(M_1) = {\rm sign}(M_2)$.
For completeness and comparison to the MRSSM, we demonstrate this
explicitly in Appendix~\ref{MSSM-positive-app}.

More interestingly, when $M_1 < 0$ and $M_2 > 0$, 
the tree-level mass difference between the lightest chargino and the 
lightest neutralino is no longer positive definite.
This is clear already at leading order in both the Higgsino
limit Eq.~(\ref{eq:mssm-split-lim-mu}) and the wino limit 
Eq.~(\ref{eq:mssm-split-lim-m2}). 
The mass difference $\Delta_+$ grows as $\tan\beta \ra 1$,
and it is also generally larger in the Higgsino limit. 
To calculate size of the splitting, especially in regions 
of the parameter space not covered by the limits given above,
we turn to numerical evaluation of the masses.

The following figures summarize the numerical results. 
First we fix $\tan \beta = 2$.  This is about the smallest value
of $\tan\beta$ that could possibly yield a lightest Higgs boson mass
above the LEP II bound \cite{Allanach:2004rh}.  In Figs.~\ref{fig:mssm3} 
and \ref{fig:mssm4} we take $\mu = 150$~GeV and
$M_2 = 150$~GeV respectively. The regions under the dotted lines
in these Figures result in $m_{\Cpmone} < 101$~GeV at tree level 
and we ignore them from further consideration.  When $\mu$ in 
Fig.~\ref{fig:mssm3} and $M_2$ in Fig.~\ref{fig:mssm4} are reduced further,
the dashed line for $M_{\Cpmone} = 101$~GeV rises.   

As before, we show contours of $\Delta_{+}$. In both of these Figures there 
are considerable regions with negative $\Delta_{+}$, demonstrating 
that the chargino can be NLSP\@.  Notice that the splitting is largest
in the Higgsino limit, up to about 5 GeV\@. 
We deliberately show a smaller range of values 
of $M_1$ and $\mu$ in Fig.~\ref{fig:mssm4} 
so that the gradient of $\Delta_{+}$ may be clearly seen.
Here $M_2$ is held fixed and we start with large $\mu$.
$\Delta_{+}$ increases as $\mu$ is decreased. The rate of increase of 
splitting peaks as $\mu$ becomes comparable to $M_2$. 

\begin{figure}[t]
\centering
\begin{minipage}[c]{0.45\linewidth}
   \centering
   \includegraphics[width=2.9in]{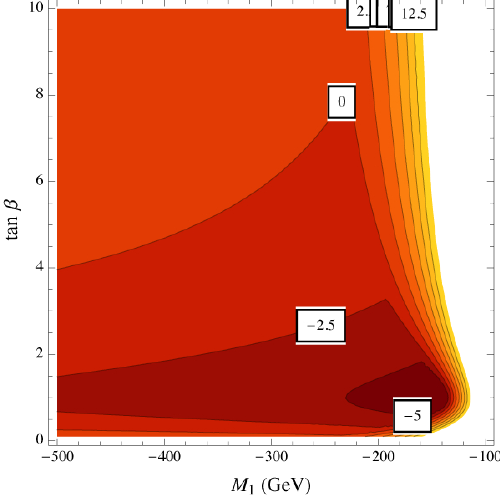} 
   \caption{Contours of $\Delta_{+}$(GeV) in the MSSM at $M_2 = 600$~GeV and 
            $\mu = 150$~GeV\@.}
   \label{fig:mssm5}
\end{minipage}
\hspace{0.5 cm}
\begin{minipage}[c]{0.45\linewidth}
   \centering
   \includegraphics[width=2.9in]{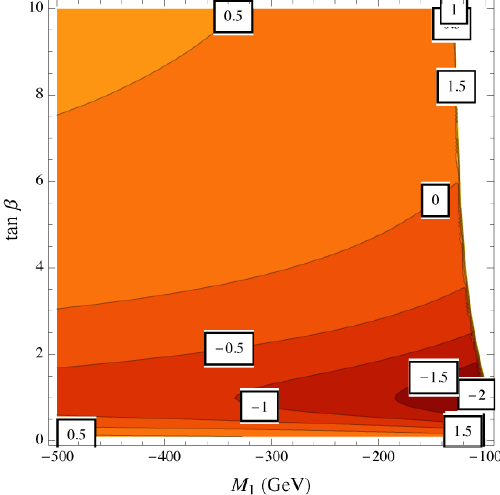}
   \caption{Contours of $\Delta_{+}$(GeV) in the MSSM at $M_2 = 150$~GeV and 
            $\mu = 250$~GeV\@.}
   \label{fig:mssm6}
\end{minipage}  
\end{figure}

We show the $\tan\beta$ dependence of the
size of the splitting in Fig.~\ref{fig:mssm5}. 
We set $M_2 = 600$~GeV and $\mu = 150$~GeV, which is a point in 
Fig.~\ref{fig:mssm3} where the splitting is the largest. We find that 
the splitting between lightest neutralino and the lightest chargino is 
maximized at $\tan\beta =1$ for a fixed value of $M_1$ and for 
$M_1 \simeq - \mu$ for a given $\tan \beta$.   For the sake of 
completeness we also evaluated the mass splitting in the
wino limit, for the parameters 
$M_2 = 150$~GeV and $\mu = 250$~GeV, in Fig.~\ref{fig:mssm6}.
Not surprisingly, the splitting is smaller than that 
in the Higgsino limit. However, the purpose of the Figure is to 
demonstrate that a wino-like chargino NLSP also occurs at 
small $\tan\beta$.

Finally, we comment on the gaugino mass hierarchy within 
a common extension of the MSSM that incorporates a gauge 
singlet $S$.  In the next-to-minimal supersymmetric standard model (NMSSM), 
the superpotential includes
\begin{equation}
\label{eq:nmssmW}
W \supset \lambda_S S H_u H_d + \frac{1}{6} k S^3 \; ,
\end{equation}
where the Higgsino mass $\mu = \lambda_S \langle S \rangle$ is generated 
after $S$ acquires an scalar expectation value.
The chargino mass matrix is unchanged while the neutralino 
mass matrix is enlarged to a $5\times5$ matrix containing 
additional parameters: the singlino mass $m_S = k \mu/\lambda_S$ 
and the singlino-Higgsino mixing $\lambda_S$. Of these parameters, 
$\lambda_S$ can always be chosen to be positive, while the sign of 
$m_S$ is arbitrary (as is the sign of $\mu$). 
Like the MSSM, we find the neutralino is heavier than the chargino whenever 
all mass parameters ($M_1, M_2$ and $m_S$) 
are positive.  However, if $m_S < 0$ the chargino can be 
the NLSP even if $M_1,M_2 >0$.  This result can be understood 
by observing that bino mixing is entirely analogous to singlino
mixing with other neutralino states, replacing the
bino mass with the singlino mass and the $g' v_{u,d}$ Higgsino
mixing terms by $\lambda_S v_{u,d}$.  Therefore, 
for the purposes of determining eigenvalues, the singlino acts 
just like the bino. As a result, the chargino can be lightest 
when either $M_1$ or $m_S$ is negative.  However, 
just like the MSSM, the chargino-neutralino splitting remains small, 
$\lesssim 5\ {\rm GeV}$.

\section{Chargino Decay}
\label{decay-sec}

Having demonstrated that the chargino can be the NLSP, we now turn 
to considering the decay of the chargino into the gravitino.  
Given that the mass splitting between the chargino NLSP and 
a neutralino next-to-next-to-lightest supersymmetric particle
(NNLSP) can be small, it may be possible for the 
neutralino NNLSP to decay directly to a gravitino.  
Hence, in the discussion below, we consider both chargino and
neutralino 2-body decays to the gravitino.  In the next section
we will compare the rates for 2-body decay of the NNLSP directly 
into a gravitino against the 3-body decay of the NNLSP into the
NLSP\@.
 
Charginos could decay to $W^{\pm}$ plus gravitino, 
or to an electrically charged ($\pm 1$, $R$-charge neutral) scalar 
plus gravitino. Similarly, neutralinos decay into a photon, $Z^0$ 
or a neutral scalar plus gravitino.  
In the MSSM, these scalars are contained 
in $H_u, H_d$, while in the MRSSM, there are extra scalars 
in $\Phi_{\tilde W}, \Phi_{\tilde B}$.  In addition
$R_u, R_d$ contain scalars, but they carry $R$-charge 2 and cannot be involved in the decay.  The form of the
decay width of a chargino (or neutralino) $\tilde{\chi}$ into a 
gravitino and a spectator particle ($X$) is~\cite{Ambrosanio:1996jn}:
\begin{equation}
\label{eq:gravwidth}
\Gamma(\tilde{\chi} \rightarrow \tilde G + X) \sim 
\frac{\kappa~m^5_{\tilde{\chi}}}{96\pi M^2_{pl}\tilde{m}^2_{3/2}}
\Big( 1 - \frac{m^2_X}{m^2_{\tilde{\chi}}} \Big)^4
\end{equation}
where $\kappa$ is an order one mixing angle. 
The decay width 
(\ref{eq:gravwidth}) is sensitive to the mass of the spectator particle 
($m_X$), and decays to heavier final states are kinematically suppressed.  
The kinematic factor makes the decays to the lightest particle possible, 
$W^{\pm}$ in the case of a chargino and $\gamma$ for a neutralino, the preferred mode. 
Additionally, the charged scalar mass matrices contains at least one extra
parameter, namely $b \ ({\rm or\ } m_{A^0})$, compared to the gaugino
mass matrices. By adjusting the additional parameter(s) we are always free
to focus on the simpler scenario where decays to charged scalars are
kinematically forbidden.  Converting the above width to a decay length
for a chargino at rest,
\begin{equation}
L = 3.4\ \kappa^{-1}\ \Big( \frac{100\ {\rm GeV}}{m_{\tilde{\chi}}} \Big)^{5}
\Big( \frac{\tilde m_{3/2}}{10\ {\rm eV}} \Big)^2
\Big( 1 - 0.6467\Big( \frac{m_{\tilde{\chi}}}{100\ {\rm GeV}} \Big)^{-2} \Big)^{-4}\ {\rm mm}\ 
\end{equation}

The characteristic decay length is proportional to $m_{3/2}^2$.  
In weakly-coupled messenger sectors, the gravitino mass can
be related to the sparticle masses.  If the mediation scale is low, 
the gravitino is typically too light to produce a visible charged track.  
With strong interactions present in the hidden sector, however, 
the gravitino mass can be significantly enhanced with respect 
to the rest of the spectrum 
\cite{Cohen:2006qc,Roy:2007nz,Murayama:2007ge},
opening up the possibility
of a chargino track.  In the following, the gravitino mass is 
taken to be a free parameter, and thus our analysis applies
regardless of the scale of messenger interactions.

Two important applications of formula (\ref{eq:gravwidth}) for
us are: the decay of the lightest chargino through
$\Cpm \rightarrow W^{\pm} \tilde{G}$, and the decay of the
lightest neutralino into $\gamma $ + gravitino, $Z + $ gravitino. 
Within the MRSSM, working in the Higgsino limit to leading order in 
$1/M_2$:
\begin{equation}
\kappa_{W\tilde G}      = \sin^2{\beta}, \quad 
\kappa_{\gamma\tilde G} = \sin^2{\theta_W}\sin^2{\beta}
                 \; \frac{M^2_W}{M^2_2}, \quad 
\kappa_{Z\tilde G}      = \sin^2{\beta} \;.
\end{equation}
Alternatively, in the wino limit we have:
\begin{equation}
\kappa_{W\tilde G}      =1, \quad \quad
\kappa_{\gamma\tilde G} = \sin^2{\theta_W}, \quad \quad
\kappa_{Z\tilde G}      = \cos^2{\theta_W} \; .
\end{equation}
to leading order in $\mathcal{O}(1/\mu)$. To lowest order 
in either limit, the $\kappa$ couplings are independent 
of $\lambda, \lambda'$.

When $M_2$ is small, the lightest neutralino is mostly the neutral wino, 
and thus the decays to $\gamma, Z$ are simply proportional to the 
photino or zino fraction of $\tilde W_3$.  On the other hand, 
when $\mu$ is small and the lightest neutralino is primarily a Higgsino, 
the coupling to the $Z$ remains order one, while the coupling to 
the photon is suppressed.

Shifting to the MSSM, in the limit $\mu \sim -M_1,$ and $|\mu|, |M_1| \ll |M_2|$ 
the lightest neutralino is a maximal mixture of Higgsino and bino, 
while the chargino is purely Higgsino. This content is reflected 
in the $\kappa $ mixing angles:
\begin{equation}
  \label{eq:kappa_mssm}
  \begin{split}
    \kappa_{W\tilde G} = & \; 1 + \sin^22\beta \; \frac{M^2_W}{M^2_2} \; , \quad \quad
    \kappa_{\gamma \tilde G} = \; \cos^2 \theta_W \Bigg[ 1 + \frac{1}{4\sqrt 2}
  \tan \theta_W \left( 1 - \sin 2\beta \right) \; \frac{M_W}{M_1} \Bigg] \; ,
                \\
   \kappa_{Z\tilde G} = & \; \frac{1}{2}\; \cos^2 \beta \; + \;
\sin^2 \theta_W \; \Bigg[ 1 \; + \; \frac{1}{4\sqrt 2}
         \tan \theta_W \Big( \frac{1 - \sin 2\beta}{\cos\beta + \sin\beta}
           \;   \Big)  \frac{M_W}{M_1}  \Bigg]  \\ 
       & \quad \quad - \;  \frac{1}{16 \sqrt 2}
      \tan \theta_W \; \Big(\frac{  2\cos\beta - 5\sin\beta + 3\sin 3\beta}
                {1 \;+\; \tan\beta} \Big) \; \frac{M_W}{M_1}  \; .
  \end{split}
\end{equation}
\section{Collider Phenomenology} 
\label{pheno-sec}

In low-scale supersymmetry models, the NLSP is typically either 
a neutralino or a charged slepton.  As all particles in a low-energy 
supersymmetry-breaking model eventually decay down to the NLSP, 
its properties such as spin, mass, charge, and decay width form the 
foundation upon which all collider studies are built. 
By demonstrating that a chargino can be the NLSPs, we are opening 
the door to an entirely new class of sparticle signatures, 
with a plethora of exciting phenomenological consequences.

Rather than study a particular exclusive process, we focus here 
on the inclusive cross sections for sparticle production with chargino NLSP\@. 
A detailed study of the optimal cuts to pick out a given sparticle spectrum 
over the background is beyond the scope of this paper; instead, our aim 
is to identify search channels and the most important SM backgrounds. 
This effort is in the same spirit as Ref.~\cite{Ambrosanio:1996jn}, 
which studied the inclusive signal with a neutralino NLSP, namely
$\gamma \gamma + \ETmiss$. 
We also comment on the possibilities for distinguishing the chargino NLSP 
from other potential NLSPs.

\subsection{Inclusive Sparticle Signal}

\begin{figure}[t]
\centering
\includegraphics[width=0.7\textwidth]{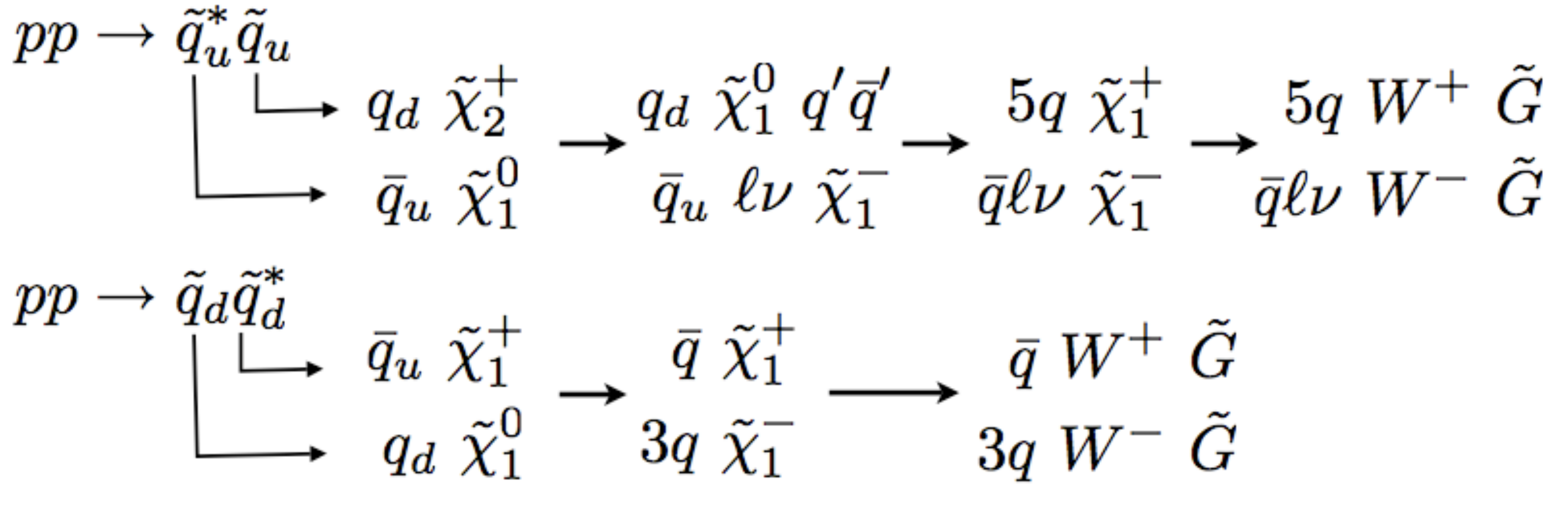}
\caption{Possible decay chains resulting from squark production.
Exactly what path is typical depends on details of the sparticle
spectrum.}
\label{chain-fig}
\end{figure}

We model the inclusive sparticle signal by the total production 
cross section for all twelve species of squarks. We do not consider 
the leptons or Higgses, and assume all squarks to be degenerate in mass. 
The inclusive squark production cross section as a function of the 
common squark mass is shown in Fig.~\ref{fig:squarkprod} below.
\begin{figure}[t]
\centering
\includegraphics[width=0.8\textwidth, height = 3.0in]{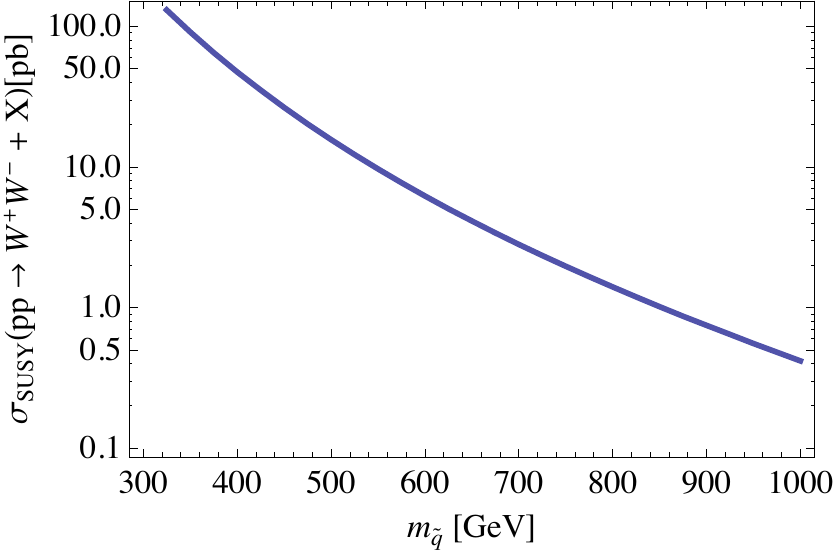}
\caption{Lowest order cross section $\sigma_{SUSY}(pp \rightarrow W^+W^- + X)$ 
at the LHC ($14\ {\rm TeV}$). We have approximated the inclusive sparticle
cross section by the QCD production cross section of all 12 squarks. 
For simplicity the squarks were taken to be degenerate, and sleptons 
were assumed to be heavier than the squarks.}
\label{fig:squarkprod}
\end{figure}
In this scenario, the squarks first decay into a quark plus chargino 
or neutralino. 
The subsequent decays of the heavy (non-NLSP) charginos/neutralinos depend on 
the details of the gaugino spectrum: if kinematically allowed, heavy 
gauginos decay to a light gaugino plus a gauge bosons,
otherwise they will decay into three-body 
decay final state containing a light gauginos plus two fermions. 
Finally, the chargino NLSPs 
each decay into $W$ + gravitino.  Thus, \emph{every} supersymmetric event 
contains at least two on-shell $W$s plus missing energy.
Using this model for sparticle production, we want to discuss the 
sparticle discovery potential via excesses in $W^+W^- + \ETmiss + X$. 
Some possible decay chains are shown in Fig.~\ref{chain-fig}.
where $\Cp_2, \N$ are the second-lightest chargino and lightest 
neutralino respectively. As we can see, in addition to on-shell $W$ pairs, 
inclusive sparticle events involve cascades containing additional 
(hard) quarks and leptons. These extra leptons/jets are especially 
important in scenarios where the chargino lives longer than 
$\sim 25\ {\rm ns}$: by the time the long-lived charginos decay 
to objects which can be triggered on, the next bunch-crossing 
will have occurred in the detector, causing the charginos to be 
associated with the wrong event. However, if there are extra 
leptons/jets present in the sparticle decays, one can trigger 
on those objects instead of on the chargino decay products. 
Then, by refining the analysis offline to search for charged tracks, 
the presence of charginos could be revealed. 

\begin{table}[h!]
\centering
\begin{tabular}{|c|c|c|}\hline
Model & Limit & $\xi^2_L + \xi^2_R$  \\ \hline
MRSSM & $\mu \ll M_2, M_1$ & $\frac{1}{2}$ \\
MRSSM & $M_2 \ll \mu, M_1$ & 2 \\ 
MSSM & $\mu \sim -M_1, M_1 \ll M_2$ & 
 $\frac{1}{4}\Big[ 1 - \frac{1}{4\sqrt 2}
  \tan \theta_W \left( \frac{1 - \sin 2\beta}{\sin\beta + \cos\beta} 
 \right) \; \frac{M_W}{M_1} \Big] $
 \\ \hline
\end{tabular}
\caption{The parameter values in different limits and models that 
determine the ratio $R_{\Gamma}$.}
\label{values-table}
\end{table}

\subsection{NNLSP Decay}

Another complication in this scenario is the possibility 
that the second-heaviest chargino and/or the lightest neutralino 
decay directly into 
gravitino + $X$ rather than decay via three-body decays.
If the heavier chargino decays directly, 
the signal will still contain two $W$ bosons plus a pair a gravitinos, 
while if a neutralino decays directly one of the $W$s is replaced by 
a hard photon or $Z$~\footnote{Neutral Higgses are also a possibility, 
which we ignore for simplicity here.}. 
The effect of the direct gravitino decays depends on the relative rates:
\begin{equation}
  \begin{split}
R_{\Gamma} = \frac{ \Gamma(\tilde{\chi}_H \rightarrow \tilde{\chi}_Lff')}
  {\Gamma(\tilde{\chi}_H \rightarrow X + \tilde G)}  & =  
    N_f \frac{4 g^4M^2_{pl}\tilde m^2_{3/2}} {5\pi^2M^4_W}
            \Big( \frac{\xi_L^2 + \xi_R^2}{\kappa}\Big)
                  \Big( \frac{\Delta}{m_{\tilde{\chi}}} \Big)^5  \\
   & = 0.556 N_f \Big( \frac{\xi_L^2 + \xi_R^2}{\kappa}\Big) \; 
       \Big(\frac{\tilde m_{3/2}}{10\ {\rm eV}} \Big)^2 \; 
        \Big(\frac{100\ {\rm GeV}}{m_{\tilde{\chi}}} \Big)^{5} \; 
           \Big( \frac{\Delta}{2\ {\rm GeV}}\Big)^5 \; , 
  \end{split}
\end{equation}
where $\Delta$ is the mass difference between the heavy 
chargino/neutralino ($\tilde{\chi}_H$) and the NLSP ($\tilde{\chi}_L$): 
$\Delta= m_{\tilde{\chi}_H} - m_{\tilde{\chi}_L}$. 
The prefactor $N_f$ is the number of fermionic degrees of freedom 
$\tilde{\chi_L}$ is kinematically allowed to decay into, and 
$\xi_L, \xi_R, (\kappa)$ are mixing angles from the chargino-neutralino-$X$ 
($\tilde{\chi}_H$-gravitino-$X$) interactions; values in the limits 
of interest are given in Table~\ref{values-table}.
We plot $R_{\Gamma}$ for the neutralino as a function 
of the mass splitting $\Delta$ and gravitino mass in 
Fig.~\ref{fig:decay-ratio} below.
\begin{figure}[t]
\centering
\includegraphics[width=0.8\textwidth, height = 3.0in]{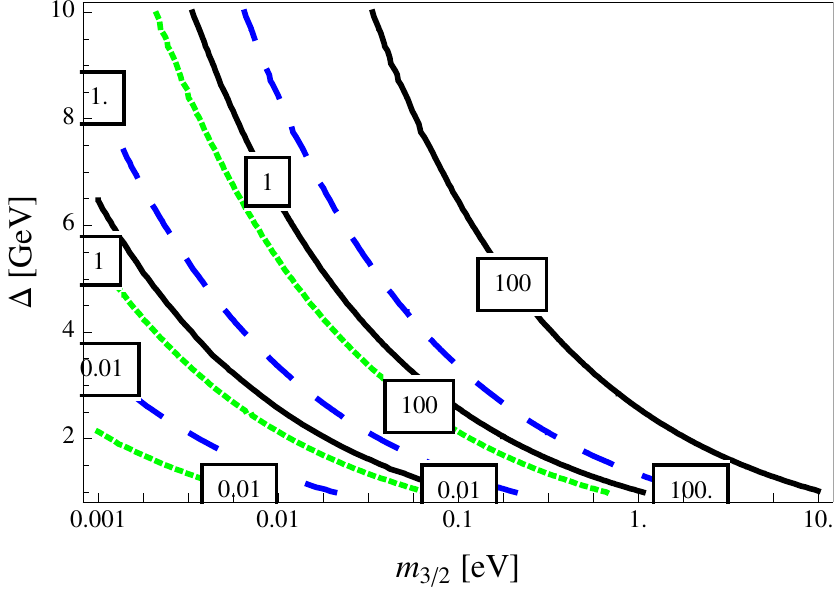}
\caption{Decay ratio $R_{\Gamma}$ as a function of the gravitino mass and the mass 
splitting $\Delta$.  $R_{\Gamma} < 1$ corresponds to NNLSPs dominantly decaying 
directly into a gravitino rather than through the NLSP\@.
The MRSSM Higgsino limit ($100\ {\rm GeV}$ NLSP, 
$\tan{\beta} = 10, M_2 = 500\ {\rm GeV}$) is the dotted line (green), the 
wino limit of the MRSSM ($100\ {\rm GeV}$ NLSP, $\tan{\beta} = 10$) is the 
dashed line (blue), and the MSSM case ($100\ {\rm GeV}$ NLSP, 
$M_1 = -200\ {\rm GeV} = -\mu,\ \tan{\beta} = 2)$ is shown as the solid line. 
In all three cases the contours indicate values, from left to right, 
of $R_{\Gamma} = 0.01, R_{\Gamma} = 1,\ {\rm and}\ R_{\Gamma} = 100$. }
\label{fig:decay-ratio}
\end{figure}
The region where direct decays are important is restricted to 
light gravitinos and small splittings. For gravitinos heavier than a 
few tenths of an eV, the three body decays are always dominant 
in either of the MRSSM limits we consider.  For the MSSM, the large 
bino component of the lightest neutralino leads to a large 
$\kappa_{\gamma \tilde G}$ and suppressed $\xi_{L,R}$, thus direct decays 
can be important  for gravitinos as heavy as $1\ {\rm eV}$.  
Neglecting direct decays of heavy gauginos to gravitinos, 
we now explore various search strategies and backgrounds to 
$W^+W^- + \ETmiss$. \\

\subsection{Search Strategies and Backgrounds}

The experimental search strategy will depend greatly on the lifetime 
of the chargino. If the chargino is long-lived but decays within 
the detector, it will leave a charged track, possibly leading to a 
displaced vertex. As mentioned above, triggering on the long-lived 
charginos is troublesome. However, provided they can be found, 
displaced vertices are great discriminant of new physics from 
background~\footnote{Lifetimes in the vicinity of $5$ ns 
would actually be ideal:  $5\ {\rm ns}$ is small enough that 
triggering/resolution is not an issue, yet long enough that 
timing information in the calorimeter could be used to differentiate 
signal from background. We thank Dirk Zerwas for bringing this point to 
our attention.}.

If the charginos decay promptly, conventional variables such as 
$H_T = \sum_{i = \ell, j}p_T$ and 
$M_{eff} = H_T + \ETmiss$~\cite{Hinchliffe:1996iu} will likely 
be the first to indicate discovery. Additionally, these variables 
will be effective handles for separating signal from background. 
The optimum value for cuts in these variables depends on the 
superpartner mass scale. 

In the case of prompt charginos, the dominant SM backgrounds in 
$W^+W^- \ETmiss + X$ are: 
$\bar t t + {\rm jets\ }$, 
single top + jets, 
$W^{+}W^{-}/W^{\pm}Z + {\rm jets}$, and 
$W^{\pm}/Z + {\rm jets}$~\footnote{By ``+jets'' we are including heavy 
quark flavors $(b,c)$ as well as light.}. Exactly which background is 
most important will depend on the strategy for $W$ pair detection. 

Irreducible SM backgrounds that result from 
$W^+ W^- \rightarrow \ell + \nu + jj$, 
contain only one true source of 
missing energy - the neutrino from the leptonic $W$ decay. 
In addition to providing less $\ETmiss$ to SM events 
(compared to two neutrinos), with a single source of missing energy
one can reconstruct the transverse mass of the $W$ ($m_{T,W}$) 
from the $p_T$ of the lepton and the $\ETmiss$. For the SM events 
the transverse mass distribution will exhibit a peak near $M_W$, 
followed by steep dropoff. Meanwhile, the $m_T$ distribution for the 
signal will fall much slower (after $M_W$) due to the excess 
$\ETmiss$ carried by the gravitinos. Therefore, by cutting on 
$m_{T,W} \gtrsim 100\ {\rm GeV}$ we can remove a large fraction 
of the background while maintaining the signal. The signal may still be 
polluted by events where two leptons are produced but one of them 
is missed by the detector. An important background which falls into 
this category is 
$\bar t t \rightarrow \ell \nu \ell' \nu' \bar b b$~\footnote{The 
$\bar t t$ and single-top backgrounds can be reduced further by 
rejecting any events with a $b$-tag, at the expense of compromising 
the stop/sbottom contribution to $(\ref{fig:squarkprod})$, while 
$W/Z + {\rm jets}$ can be reduced by requiring two jets to reconstruct a $W$.}.

A second channel to search for $W$ pairs, $\ell\nu \ell' \nu'$, has the 
advantage of an additional lepton, which greatly reduces the 
$W^{\pm}/Z ~+$ jets background. Unfortunately, when both $W$'s decay 
leptonically, the SM backgrounds have higher missing energy, 
so the efficiency of $\ETmiss$ cuts will be reduced. 
The $p_T$ and $\eta$ of the two leptons remain useful variables for 
discriminating signal from background.

The third possibility is for both $W$s to decay hadronically. In that case,
the signal is 4+~jets + $\ETmiss$ , with four of the jets breaking up 
into two pairs, each pair reconstructing to a $W$.  
Although reconstructing $W$s can be 
difficult with realistic jet resolution, the background can be 
greatly suppressed by imposing hard cuts ($100\ {\rm GeV}$ or more) 
on the $p_T$ of the reconstructed $W$s and the missing energy. 
However, unless there is an additional lepton in these events from an earlier 
cascade decay, multijet QCD becomes an 
important and difficult background.  
Further detailed study, in all of the channels mentioned 
above as well as for a more diverse spectrum are needed.

\subsection{Confounding NLSPs}

If the chargino NLSP is long-lived, a charged track will be visible 
in the detector.  Depending on the exact lifetime of the chargino, 
the charged track will either exit the detector or end in a 
displaced vertex.  Either way, a visible track allows us to rule out 
the majority of neutralino NLSP scenarios.  However, distinguishing 
a chargino NLSP from a charged slepton NLSP requires more careful
study.

As sleptons are much heavier their decay products, the lepton coming 
from the displaced vertex will tend to be collinear with the 
slepton track. On the other hand, because $W$s  are so massive they 
will emerge from chargino decays without significant 
boosting; therefore chargino tracks will demonstrate a distinct kink feature.  
Additionally, $W$ bosons decay to jets and (democratically) into all 
three leptons - 
therefore chargino decays will lead to $e, \mu,\tau$ each in equal amounts, 
up to reconstruction and detector efficiency effects. 
Selectron and smuon NLSPs also decay to lepton plus $\ETmiss$, 
but their final state is a particular lepton flavor 
(for at least approximately flavor-diagonal slepton masses).
By simply counting the number and type of leptons in a data sample, 
one should easily be able to distinguish selectrons/smuons from charginos.  
Stau NLSPs are somewhat trickier because taus decay democratically 
to electrons and muons; however, provided adequate tau-tagging capability 
at the LHC, staus should also be easily distinguishable from charginos. 

Short-lived charginos are somewhat trickier. In principle, the 
identity of the particle produced along with the gravitino can tell 
us something about the NLSP; for neutralinos this spectator particle 
is likely a photon,  for a chargino it will be a $W$, and for sleptons 
the spectator is a lepton. Additionally, the maximum energy achieved 
by the spectator depends on the LSP mass, allowing one to distinguish 
WIMP LSP scenarios from gravitino LSP scenarios~\cite{Dimopoulos:1996vz}. 
In practice, the success of spectator-identification will depend on 
the details of the sparticle spectrum, and it is easy to dream up 
tricky scenarios. 

To distinguish the chargino NLSP limit of the (N)MSSM from the MRSSM 
additional observables are needed.  One handle is that the hallmark 
same-sign-lepton signal will be absent~\cite{Chacko:2004mi}
whenever the gauginos have a Dirac mass. A second, though less robust 
discriminant is the mass difference between the lightest neutralino 
and the lightest chargino. In the MSSM the NLSP chargino occurs when 
several gauginos are nearly degenerate, thus it is difficult to arrange 
mass differences (even with large radiative corrections) greater 
than $\sim 10\ {\rm GeV}$. In the MRSSM large neutralino-chargino 
mass differences are easier to accommodate, especially if 
$\lambda, \lambda' \ne 0$.

\section{Conclusions}
\label{conclusions-sec}

In this paper we have studied a new signal of supersymmetry that results
when a chargino is the NLSP and the gravitino is LSP\@. 
A necessary condition for this scenario to occur is that
a chargino must be the lightest gaugino. We have found
\begin{itemize}
\item The chargino can be the lightest gaugino in a wide range of parameter
space when neutralinos are Dirac fermions, such as in the MRSSM\@. 
\item In the MSSM a chargino can be the NLSP essentially only in the case
${\rm sign}(M_1) \not= {\rm sign}(M_2) = {\rm sign}(\mu)$. 
\item There is qualitative difference between the generated 
gaugino mass hierarchies depending on whether 
the neutralinos are Dirac fermions or Majorana fermions. In the MSSM,
the splitting is large when $\tan\beta$ is small, as opposed to the 
case of Dirac gauginos when the splitting is maximized for large $\tan\beta$. 
\item In addition we also observe quantitative difference between the two cases.
In the MSSM the splitting is small ($\lesssim \mathcal{O}(5\ \rm{GeV})$) 
and in the MRSSM the splitting can be as big as 
$\mathcal{O}(30\ \rm{GeV})$ at tree level.
\end{itemize}
Given a light gravitino, a chargino NLSP will decay
into a gravitino and an on-shell $W$. Summarizing the phenomenology:
\begin{itemize}
\item If the gravitino mass is far larger than 
${\cal O}(100 \; {\rm eV})$, a chargino produced at a collider
will escape the detector leaving a charged track.
\item If the gravitino mass is of order 
${\cal O}(1-100 \; {\rm eV})$, a chargino produced at a collider
can have a displaced vertex and/or track, resulting in a
decay well away from the interaction vertex.
\item If the gravitino mass is far smaller than
${\cal O}(1 \; {\rm eV})$, a chargino produced at a collider
will decay promptly into a $W$ and a gravitino.
\end{itemize}
In the first two cases, the charged track provides a great way to discriminate
signal from background.
Distinguishing a chargino NLSP from a slepton NLSP requires exploiting the
smaller boost
and flavor-democratic decay of the $W$.
When the chargino decays promptly, the pair of $W$s in each event provide a
striking signature
of sparticle production. Standard methods may suffice to reduce background and
extract the signal,
however additional benefit could be achieved by taking advantage of the
characteristic
$WW + \slashchar{E}_T + X$ final state.  Dedicated studies of exclusive
sparticle production
may provide more promising opportunities to discover supersymmetry with a
chargino NLSP\@.

\begin{acknowledgments}

We thank S.~Thomas, N.~Weiner and D.~Zerwas for helpful conversations.
GDK and AM thank the Aspen Center for Physics for
hospitality where part of this work was completed.
GDK acknowledges the support of the KITP Santa Barbara, where some 
of this work was performed, and the support in part by the 
National Science Foundation under Grant No. PHY05-51164.
This work was supported in part by the DOE under contracts
DE-FG02-96ER40969 (GDK, TSR) and DE-FG02-92ER40704 (AM).

\end{acknowledgments}
\begin{appendix}

\section{The Fox-Nelson-Weiner model with Dirac gaugino masses}
\label{supersoft-app}

The FNW model \cite{Fox:2002bu} contains the same content
as the MRSSM but \emph{without} the $R_u,R_d$ fields.
Gaugino masses arise exclusively from the ``supersoft'' operator,
Eq.~(\ref{dirac-eq}).
The Higgs sector in this model, however, has an ordinary $\mu$-term
in place of Eq.~(\ref{eq:mu-term}).  Just like the MSSM, the
$\mu$-term marries the Higgsinos $\tilde{H}_u$ and $\tilde{H}_d$ 
with each other, and after expanding around the nonzero vevs of 
the two Higgses, one finds mass terms between the gauginos and 
the Higgsinos. The resulting neutralino mass terms are given by
\begin{equation}
  \mathcal{L} \supset \frac{1}{2} \; N_0^\text{T} M_n N_0 \; ,
\end{equation}
where 
\begin{equation}
N_0 =
 \left[ \begin{array}{c}  
     \tilde{W}_3 \\  \tilde{B} \\ \tilde{H}_d^0 \\
     \psi^0_{\tilde{W}} \\  \psi_{\tilde{B}} \\  \tilde{H}_u^0 
      \end{array} \right]   \quad  \text{and}  \quad 
M_n =
  \left[ \begin{array}{cccccc}  
      0 & 0 &    g  v_d / \sqrt{2} & M_2 & 0 &  -  g  v_u/\sqrt{2}  \\ 
      0 & 0 &  -  g' v_d / \sqrt{2} & 0  & M_1 &  g' v_u/\sqrt{2}  \\ 
        g  v_d / \sqrt{2} & - g' v_d / \sqrt{2} &  0 & 0 & 0   & \mu  \\ 
      M_2 &  0  & 0 & 0 & 0 & 0 \\
      0   & M_1 & 0 & 0 & 0 & 0 \\
      - g  v_u / \sqrt{2} &   g' v_u / \sqrt{2} & \mu & 0 & 0   & 0  
       \end{array} \right]  \; .
\end{equation}
The fields $\psi_i$ are the fermions in the supermultiplets $\Phi_i$. In 
particular, $\psi_{\tilde{W}}$ is a $SU(2)_W$ triplet containing
a neutral $\psi^0_{\tilde{W}}$ and as well as charged components
$\psi^{\pm}_{\tilde{W}} \equiv \frac{1}{\sqrt{2}} 
\left( \psi^{(1)}_{\tilde{W}} \pm i \psi^{(2)}_{\tilde{W}} \right)$.
The chargino mass terms in this model are given by:
\begin{equation}
 \mathcal{L} \supset
    \left[ \begin{array}{ccc}
     \tilde{W}^{+} &  \psi^{+}_{\tilde{W}} & \tilde{H}_u^{+}       
     \end{array} \right]  
       \left[ \begin{array}{ccc}
        M_2 & 0 &   g  v_d   \\   
        0  & M_2 &  0 \\ 
        0 &  g v_u   & \mu 
        \end{array} \right] 
      \left[ \begin{array}{c}  
          \psi^{-}_{\tilde{W}} \\  \tilde{W}^{-} \\  \tilde{H}_d^{-} 
      \end{array} \right]  \; .
\end{equation}

Similar to the discussion of the MRSSM, in the limit of large $\tan\beta$ 
(where $v_d \rightarrow 0$) the neutralino mass matrix simplifies
drastically.  A linear combination of $U(1)_R$ under which 
the gauginos are charged and $U(1)_Y$ under which the Higgsinos 
are charged is preserved and thus neutralinos become Dirac fermions 
with the following mass matrix:
\begin{equation}
 \mathcal{L} \supset
    \left[ \begin{array}{ccc}
     \tilde{W}_3 &  \tilde{B} & \tilde{H}_d^0       
     \end{array} \right]  
       \left[ \begin{array}{ccc}
        M_2 & 0 &  - g  v_u/\sqrt{2}  \\   
        0  & M_1 &   g' v_u/\sqrt{2}  \\ 
        0 & 0   & \mu 
        \end{array} \right] 
      \left[ \begin{array}{c}  
          \psi^{0}_{\tilde{W}} \\  \psi_{\tilde{B}} \\  \tilde{H}_u^0 
      \end{array} \right]  \; .
\label{FNW-neut-simp-eq}
\end{equation}
The chargino mass matrix also simplifies:
\begin{equation}
 \mathcal{L} \supset
    \left[ \begin{array}{ccc}
     \tilde{W}^{+} &  \psi^{+}_{\tilde{W}} & \tilde{H}_u^{+}       
     \end{array} \right]  
       \left[ \begin{array}{ccc}
        M_2 & 0 &   0  \\   
        0  & M_2 &  0 \\ 
        0 &  g v_u   & \mu 
        \end{array} \right] 
      \left[ \begin{array}{c}  
          \psi^{-}_{\tilde{W}} \\  \tilde{W}^{-} \\  \tilde{H}_d^{-} 
      \end{array} \right]  \; .
\label{FNW-char-simp-eq}
\end{equation}

Despite $U(1)_R$ broken by the Higgs sector in the model,
it is easy to see that a chargino can be the NLSP\@.
In the large $\tan\beta$ limit, the mass matrices in 
Eqs.~(\ref{FNW-neut-simp-eq}) and (\ref{FNW-char-simp-eq}) 
are identical to what was obtained in the MRSSM
in the case when the two Higgsinos have equal masses
(i.e.\ $\mu_u = \mu_d = \mu$).  Similarly, when
$\mu_d \gg M_1, M_2, \mu_u = \mu$ and $\tan\beta$ is large,
the relevant part of the mass matrices in the MRSSM in 
Eqs.~(\ref{eq:mrssm-nut-mass}) and (\ref{eq:mrssm-chr-mass})
become identical to Eqs.~(\ref{FNW-neut-simp-eq}) and 
(\ref{FNW-char-simp-eq}) respectively.

\section{Verifying the Form of the MRSSM Gaugino Mass Matrices}
\label{checks-app}

One important verification results by setting all supersymmetry 
breaking parameters as well as all $\lambda$ parameters to zero. 
In this limit, the masses of the neutralinos and charginos arise 
due to the vev of the Higgses and their kinetic terms. 
The kinetic terms of the Higgses are obviously independent
of $R$-symmetry.
This implies that the masses of the heavy charginos and neutralinos 
should be identical whether in the MSSM or in the MRSSM\@.  
This requirement verifies that the elements in the mass matrices 
proportional to gauge couplings have the correct form.

In order to verify the rest of the couplings we expanded the 
corresponding terms in the Lagrangian in terms of electromagnetic 
eigenstates in the place of weak eigenstates. The numerical factors 
between the couplings of the charge neutral states and the charged 
states now determine various elements in the mass matrices. 
\begin{equation}
  \begin{split}
   H_u R_u   \quad \quad & \rightarrow    \quad \quad 
         H_u^0 R_u^0 \; + \; H_u^{+} R_u^{-} \\
   H_u \Phi_{\tilde{W}} R_u    \quad \quad  & \subset   \quad \quad  
       \frac{1}{\sqrt{2}} H_u^0 \left( \frac{1}{\sqrt{2}} \Phi^0_{\tilde{W}}  R_u^0 \;  + 
         \Phi^{+}_{\tilde{W}}  R_u^{-} \right)   \\
   W \Phi_{\tilde{W}}   \quad \quad & \rightarrow    \quad \quad  
        W^{+} \Phi^{-}_{\tilde{W}} +  W^{-} \Phi^{+}_{\tilde{W}}
        + W_3 \Phi^{0}_{\tilde{W}} \; ,
  \end{split}
\end{equation}
where the charge eigenstates are defined the usual way.

\section{Chargino and Neutralino Masses in the MSSM with
${\rm sign}(M_1) = {\rm sign}(M_2)$}
\label{MSSM-positive-app}
\begin{figure}[t]
\centering
\begin{minipage}[c]{0.45\linewidth}
   \centering
   \includegraphics[width = 2.9 in]{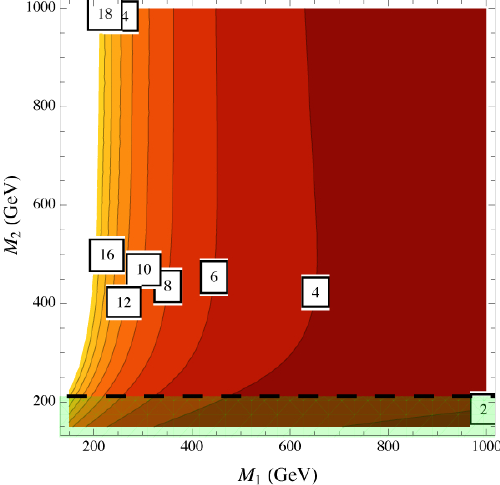} 
   \caption{Contours of $\Delta_{+}$ (GeV) in the MSSM at $\tan\beta=2$ and 
            $\mu = 150$~GeV\@.}
   \label{fig:mssm1}
\end{minipage}
\hspace{0.5 cm}
\begin{minipage}[c]{0.45\linewidth}
   \centering
   \includegraphics[width = 2.9 in]{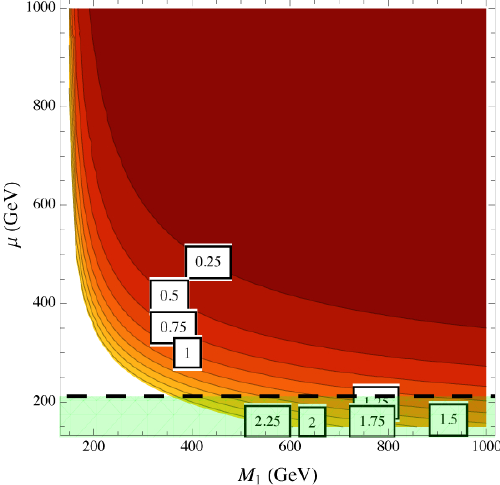}
   \caption{Contours of $\Delta_{+}$ (GeV) in the MSSM at $\tan\beta=2$ and 
            $M_2 = 150$~GeV\@.}
   \label{fig:mssm2}
\end{minipage}  
\end{figure}

It is illuminating to numerically explore a broader range of the 
MSSM parameter space. In Figs.~\ref{fig:mssm1} and \ref{fig:mssm2}  we plot the
contours of $\Delta_{+}$, the difference of 
the lightest chargino mass to the lightest neutralino mass. 
Fig.~\ref{fig:mssm1} explores the mostly-Higgsino limit and was generated by holding 
$\mu = 150$~GeV and $\tan\beta = 2$.
Similarly, Fig.~\ref{fig:mssm2} explores the 
mostly-wino limit and was generated holding 
$M_2 = 150$~GeV and $\tan\beta = 2$.  In each case, the Figures 
clearly show  $\Delta_+ >0$ throughout the parameter space, 
and thus the neutralino is always lighter than the chargino. 
This confirms the results of the limiting cases 
Eq.~(\ref{eq:mssm-split-lim-mu}) and Eq.~(\ref{eq:mssm-split-lim-m2}).  
Moreover, these Figures also demonstrate that a neutralino remains 
the lightest gaugino even when all the parameters in the mass matrices 
are of the same order.  
The regions under the dashed lines
in these Figures result in $m_{\Cpmone} < 101$~GeV at tree level.
When $\mu$ in 
Fig.~\ref{fig:mssm1} and $M_2$ in Fig.~\ref{fig:mssm2} are reduced further,
the dashed line for $M_{\Cpmone} = 101$~GeV rises. For example setting 
$\mu = 120$~GeV excludes the entire region below $M_2 = 450$~GeV\@.  

The numerical results carry additional insight into the 
chargino/neutralino system.  A common feature of 
Fig.~\ref{fig:mssm1} and Fig.~\ref{fig:mssm2} is that the 
contours decrease as one moves to larger values of both the 
$x-$ and $y-$coordinates (to the top-right corner). 
In Fig.~\ref{fig:mssm1} this can be understood by noting 
that if both $M_1,M_2$ are heavy, the gauginos associated 
with them (the bino and winos) can be integrated out;
the low energy effective theory is left with just Higgsinos, 
which contain a pair of degenerate neutralinos and charginos,
{\it i.e.} $\Delta_{+} \rightarrow 0$ in the decoupling limit.  
Exactly the same arguments go through for the wino limit 
(with increasing $\mu$ and $M_1$),
shown in Fig.~\ref{fig:mssm2} where the decoupling is
much more rapid.  
This provides additional verification of our analytical results  
in Eq.~(\ref{eq:mssm-split-lim-mu})
where we saw that the splitting was more suppressed in the
wino case.

\end{appendix}


\begin{thebibliography}{10}
\bibitem{othersignnote}
We have also found that the chargino can be the NLSP at tree-level
when ${\rm sign}(M_1) \not= {\rm sign}(M_2) = -\mu$.  In this 
region we find the mass difference is smaller than 1\% of the
NLSP mass.  It is not obvious that this region remains after
radiative corrections are taken into account, and thus we do 
not consider this region further.
\end{thebibliography}
\end{document}